# Spatial competition and the dynamics of rarity in a temporally varying environment

Lauren O'Malley[1], G. Korniss[1], Sai Satya Praveen Mungara[2] and Thomas Caraco[3]

[1] *Department of Physics, Applied Physics, and Astronomy, Rensselaer Polytechnic Institute, 110 8th Street, Troy, NY 12180-3590, USA*
[2] *Department of Computer Science, University at Albany, Albany, NY 12222, USA*
[3] *Department of Biological Sciences, University at Albany, Albany, NY 12222, USA*

## ABSTRACT

**Questions**: How does temporal variation in competitive advantage affect advance from rarity and species abundances in an individual-based ecology? In particular, how does the difference between the timescale of competitive invasion and the timescale of environmental periodicity interact with the spatial clustering underlying invasion to influence global population dynamics?

**Features of Model**: We assume that two species compete preemptively for space in a two-dimensional environment. We categorize invasion as nucleation and growth of one cluster of the rare species, or as nucleation of many clusters. Simulation of a constant environment identifies the characteristic timescale for a competitively superior species to invade and numerically dominate a resident species.

**Manipulation of Key Variables**: Given an endogenous timescale set by invasion in a constant environment, we introduced periodic temporal variation in competitive superiority by alternating the species' propagation rates. We set the half-period of the environment much less than, roughly equal to, and much greater than the endogenous timescale. By manipulating habitat size and introduction rate, we simulated environments where successful invasion proceeds through growth of many spatial clusters, and where invasion can occur only as a single-cluster process.

**Conclusions**: In the multi-cluster invasion regime, rapid environmental variation produced spatial mixing of the species and non-equilibrium coexistence. The dynamics' dominant response effectively averaged environmental fluctuation, so that each species could avoid competitive exclusion. Increasing the environment's half-period to match the population-dynamic timescale let the (initially) more abundant resident repeatedly repel the invader. Periodic transition in propagation-rate advantage rarely interrupted the exclusion process when the more abundant species had competitive advantage. However, at infrequent and randomly occurring times, the rare species could invade and reverse the density pattern by rapidly eroding the resident's preemption of space.

In the single-cluster invasion regime, environmental variation occurring faster than the population-dynamic timescale prohibited successful invasion; the first species to reach its stationary density (calculated for a constant environment) continued to repel the other during long simulations. When the endogenous and exogenous timescales matched, the species randomly reversed roles of resident and invader; the waiting times for reversal of abundances indicate stochastic resonance. For both invasion regimes, environmental fluctuation occurring much slower than the endogenous dynamics produced symmetric limit cycles, alternations of the constant-environment pattern.

Correspondence: Thomas Caraco, Department of Biological Sciences, University at Albany, Albany, New York 12222, USA. e-mail: caraco@albany.edu

# INTRODUCTION

Ecologists recognize that local dispersal induces spatially correlated population densities (Ellner *et al.*, 1998; Wilson, 1996; 1998). These correlations can govern interaction frequencies and, consequently, impact the global dynamics of competing species (Tilman and Kareiva, 1997; Chesson, 2000; Dieckmann *et al.*, 2000). Models for spatially detailed competition generally predict that conditions discriminating coexistence from exclusion can depend on the degree of local structure (*e.g.*, Bolker and Pacala, 1999; Mágori *et al.*, 2005; Caraco *et al.*, 2006). More subtly, but no less significantly for our understanding of competition, interactions structured by local dispersal often increase the characteristic time scale of population dynamics well beyond that of mean-field models (Hurtt and Pacala, 1995; Lehman and Tilman, 1997; O'Malley *et al.*, 2006a). For example, the time elapsing between introduction of a superior competitor and displacement of an ecologically less efficient, resident species can be far longer when individuals interact only at the neighborhood scale, compared to a well-mixed dynamics (Gandhi *et al.*, 1999; Korniss and Caraco, 2005; O'Malley *et al.*, 2005).

Ecologists further recognize that environmental variation between generations can strongly influence competitors' dynamics (Chesson 1990; Ripa and Ives, 2003; Descamps-Julien and Gonzalez, 2005; Adler and Drake, 2008). Environmental fluctuations might reduce densities sufficiently that the chance of extinction increases. However, temporal variation, at some periodicities, might help prevent competitive exclusion (Hutchinson, 1961; Caswell and Cohen, 1995; Spencer *et al.*, 2007; D'Odorico *et al.*, 2008). In particular, asynchrony between competing species' rates of growth in a fluctuating environment may promote coexistence via temporal niche differences (Chesson and Huntly, 1997; Snyder, 2007).

Only rarely have ecologists combined analyses of spatially heterogeneous populations with environmentally induced demographic fluctuation (Chesson, 1990; Holt and Barfield, 2003; Schoolmaster and Snyder, 2007). Our study takes an individual-based approach to this interaction; we model preemptive competition in an environment subject to temporal variation. Given an exogenous process that periodically reverses competitive superiority between two locally-dispersing species, we ask how this individual-level variation affects the competitors' global dynamics. The few previous individual-based ecological models with temporal variation (*e.g.* Holt *et al.*, 2004) ordinarily consider a single population's growth only. But recently Chan *et al.* (2009) studied competitive coexistence with a multitype contact process subject to seasonal variation; see Discussion.

We categorize our results according to the difference between two fundamental timescales, in each of two competitive-invasion regimes. The first temporal scale is the expected time $\langle \tau \rangle$ required for a superior competitor to invade a resident species and advance to numerical dominance in a constant environment (endogenous timescale). The second temporal scale is the half-period of the environmentally induced demographic-rate variation $t_{1/2}$, the time elapsing between changes in competitive rank (exogenous timescale). Invasion regimes distinguish between a successful invader growing as a single spatial cluster and invasive growth distributed among many clusters (Korniss and Caraco, 2005). Similar analyses of interacting timescales in physical systems (e.g., Korniss *et al.*, 2000; Buendia and Rikvold, 2008) suggest a set of metrics, and offer predictions paralleling our results on the ecology of spatial growth; see Discussion.

We organize the paper as follows. First, we briefly summarize an individual-based model for preemptive competition, to compare the fundamental timescales of single-cluster and multi-cluster invasion in a constant environment. Next, we introduce periodic variation in reproductive rates, and interpret a series of model simulations. We employ some simple metrics, functions of the difference between the competitors' densities, revealing the range of dynamic complexity emerging from the interaction of the time-scale comparison with invasion regime. Finally, we summarize the dynamics and place our work and in context.

## NUCLEATION AND ECOLOGICAL INVASION

Before addressing temporal variation, we recall our constant-environment model, which draws on nucleation theory for clustered growth, sometimes referred to as KJMA theory - for Kolmogorov (1937), Johnson and Mehl (1939) and Avrami (1940, 1941). Nucleation theory predicts timescales associated with spatial growth in multi-cluster systems; recent applications of the theory span the physical (Rikvold *et al.*, 1994; Ramos *et al.*, 1999; Korniss *et al.* 2002) and biological sciences (Gandhi *et al.*, 1999; Herrick *et al.*, 2002; Jun *et al.*, 2004; O'Malley *et al.*, 2005; Zhang and Bechhoefer, 2006).

Suppose that a competitively inferior species advances to its self-regulated, equilibrium global density. Thereafter, individuals of a competitively superior species are introduced stochastically at rate much smaller than the two species' birth and death rates; introduction occurs rarely, but repeatedly. Given a sufficiently high initial density of the competitively inferior resident, and preemptive competition (Platt and Weis, 1985; Connolly and Muko, 2003; Yurkonis and Meiners, 2004; Rácz and Karsai, 2006), the resident can resist invasion for a long time before declining (perhaps suddenly) toward competitive exclusion. A dynamics where rare, random introduction events combine with strongly clustered growth of the invader – features of many plant and animal invasions (D'Antonio, 1993; Holway, 1988; Herben *et al.*, 2000) – cannot be captured faithfully by either mean-field models (homogeneous mixing) or deterministic partial differential equations (Moro, 2001; Antonovics *et al.*, 2006). Nucleation theory, however, provides a powerful framework to model growth of a locally dispersing invader.

### Individual-based model of spatial competition

To model preemptive competition, we consider an $L \times L$ lattice with periodic boundaries. Each site represents the resources required to sustain a single individual. The local occupation number at site $\mathbf{x}$ is $n_i(\mathbf{x}) = 0$, 1 with $i = 1, 2$, representing the number of resident and invader individuals, respectively. An empty site may be occupied by species $i$ through propagation from a neighboring site at rate $\alpha_i \eta_i(\mathbf{x})$, where $\alpha_i$ is the individual-level propagation rate for species $i$, and $\eta_i(\mathbf{x}) = \left( \frac{1}{\delta} \right) \sum_{\mathbf{x}' \in \mathbf{nn}(\mathbf{x})} n_i(\mathbf{x}')$ is the density of species $i$ in the neighborhood about site $\mathbf{x}$. $\mathbf{nn}(\mathbf{x})$ is the set of the $\delta$ nearest neighbors of site $\mathbf{x}$; in this study we fix $\delta = 4$. Each species also may occupy an empty site through immigration from outside the environment; each species has introduction rate $\beta$ per open site. Introduction and local propagation occur independently, so that species $i$ occupies an open site at total probabilistic rate $\beta_i + \alpha_i \eta_i$, for $i = 1, 2$. An occupied site opens through density-independent mortality of the individual; here each species has mortality rate $\mu$.

For constant environments, we take $\beta << \mu < \alpha_1 < \alpha_2$; introduction is rare, and invaders have a per-individual reproductive advantage. In the simulations, we track the time-dependent global density of each species, $\rho_i(t) = \left( \frac{1}{L^2} \right) \sum_{\mathbf{x}} n_i(\mathbf{x}, t); i = 1, 2$. We define the resident's lifetime $\tau$ as the first passage time of the resident's global density to ½ its initial (hence, quasi-equilibrium) density $\rho_1^*$. For given parameters, we take the resident's mean lifetime $\langle \tau \rangle$, invasion time for simplicity, as the characteristic time scale of the population dynamics. The ½ is arbitrary; we choose it since the invader's global density $\rho_2(t)$ should just exceed the resident's density when $\rho_1(t) = \frac{\rho_1^*}{2}$. The term "invasion time" fits the constant-environment dynamics, since the superior invader always becomes common (though the elapsed time varies randomly). Since we introduce individuals of each species, competitive exclusion does not imply true extinction. Rather, we equate exclusion with a small global density $O(\beta)$, where the common introduction rate satisfies $10^{-8} \leq \beta \leq 10^{-4}$.

As simulation proceeds, individual invaders occasionally appear interspersed among the initially common residents. An invader lacking access to empty, neighboring sites may die before propagating. If a site opens in the local neighborhood, the invader may colonize it. However, the resident's greater local density may compensate for its lower individual-level propagation rate, so that the resident species has the greater chance of colonizing the empty site. Consequently, most small invader clusters shrink and disappear. Residents, though weaker competitors, can avert exclusion for extended periods, since preemptive competition constrains invader growth. Invaders can succeed only after they generate a cluster sufficiently large (the radius exceeds a critical radius $R_{crit}$) that it tends to grow at its periphery (Yasi *et al.*, 2006; Allstadt *et al.*, 2007). We say a nucleation event occurs when a cluster's radius first reaches $R_{crit}$.

**Single-cluster and multi-cluster invasion**

For given parameters, let $D(\alpha_1, \alpha_2, \mu, \beta)$ represent the average distance separating invader clusters in a hypothetically infinitely large environment. If the size of the actual environment is sufficiently small, $L \ll D$, then invasion almost always proceeds through the growth of a single cluster (SC) invasion. However, when $L \gg D$, invasion involves many invader clusters [multi-cluster (MC) invasion]. Furthermore, suppose that we fix the linear habitat size $L$, as well as other parameters - except the introduction rate Then there exists a characteristic value of the introduction rate $\beta$ (now governing $D$) such that MC invasion crosses over to the SC pattern for levels of $\beta$ less than the characteristic value (O'Malley *et al.*, 2006a).

We can distinguish SC and MC invasion, and identify ecological implications, within the framework of homogeneous nucleation (Korniss and Caraco, 2005). We outline the essentials in Appendix 1; here we specify how the endogenous time scale differs between modes of invasion.

For SC invasion, the lengthy waiting time until the first successful cluster nucleates dominates the lifetime $\tau$ of the competitively inferior resident. Differences between the competitors' demographic rates govern the time required for the cluster to grow, but growth time is short compared to the exponentially distributed waiting time for nucleation (O'Malley *et al.*, 2006a). Hence if $t_n$ is the random waiting time until nucleation occurs (the first invader cluster with a radius as large as $R_{crit}$), then $\langle \tau \rangle \approx \langle t_n \rangle$. From Appendix 1, $\langle t_n \rangle \approx \left( L^2 \beta \right)^{-1}$. Therefore, the characteristic timescale, $\langle \tau \rangle$, varies inversely with both habitat area and the introduction rate (O'Malley *et al.*, 2005). To compare invasion modes, we can say the endogenous timescale $\langle \tau \rangle$ scales with $\beta^{-1}$ in the SC regime. Single-cluster invasion predicts that an invader maintains a low mean density, perhaps below a detection threshold, for an uncertain length of time – which can be quite long. Then, stochastically, a cluster grows beyond critical radius, and the invader drives the resident toward competitive exclusion.

For large environments (or large introduction rates), $L \gg D$; the environment's size exceeds the typical distance between invader clusters. Consequently, many randomly nucleated, expanding clusters drive the resident's decline. In the limit of a large environment, we can approximate the time-dependent global densities by Avrami's law (see Appendix 1) which provides a functional form for global densities during MC invasion. Note that for MC invasion, $\langle \tau \rangle$ is essentially independent of the habitat size (Korniss and Caraco, 2005). Finally, for yet larger $\beta$, invader clusters coalesce immediately, and nucleation theory breaks down; in fact, homogeneous mean-field equations may apply as clusters mix (O'Malley *et al.*, 2006b).

Invasion time $\langle\tau\rangle$, our population-dynamic timescale, depends on introduction rate $\beta$, and this dependence differs between invasion regimes. For MC invasion, the timescale of the invader's advance to numerical superiority in simulation scales as $\langle\tau\rangle \sim \beta^{-0.30\pm0.015}$, close to $\beta^{-\frac{1}{3}}$, the value predicted by Avrami's law (O'Malley *et al*., 2006a). However, for any finite environment with linear size $L$, a sufficiently small $\beta$, such that $D(\alpha_1, \alpha_2, \mu, \beta) \sim \beta^{-\frac{1}{3}} >> L$, implies SC invasion. In the SC regime, the average lifetime (and its standard deviation) scales as $\langle\tau\rangle \sim \beta^{-1}$, reflecting the Poisson nature of an invader cluster's nucleation; see Appendix 1. For extended analysis of invasion time *versus* invasion regime, see O'Malley *et al*. (2005).

## TEMPORAL VARIATION AND SPATIAL COMPETITION

Most communities experience some level of temporal variation (Descamps-Julien and Gonzalez, 2005; Greenman and Norman, 2007; Adler and Drake, 2008; Loreau and de Mazancourt, 2008), and these environmental fluctuations may impact ecological and evolutionary processes (Caswell and Cohen, 1995; Ives, 1995; Chesson, 2000; Neubert *et al*., 2000; Travis *et al*., 2005; Altizer *et al*., 2006; Spencer *et al*., 2007). Given our model's behavior in a constant environment, we ask how the dynamics responds to periodic temporal variation. The distinction between single and multi-cluster invasion proves useful in a periodically varying environment (Korniss *et al*., 2002).

### Time-dependent propagation rates

To study a symmetric demographic response to environmental fluctuation, we modified our individual-based model by setting $\alpha_1(t) = \alpha + \Delta(t)$, and $\alpha_2(t) = \alpha - \Delta(t)$. We report results where $\Delta(t)$ is a square-wave with amplitude $\Delta$ and half-period $t_{1/2}$; see Buceta *et al*. (2003). Since each species has the same individual mortality rate (we fix $\mu = 0.1$ throughout this paper), the environment alternately favors one, then the other species via the propagation-rate difference. Here we let $\alpha = 0.75$ and $\Delta = 0.05$, so that $[\alpha_1(t), \alpha_2(t)]$ alternates between [0.8, 0.7] and [0.7, 0.8]. We selected the invasion regime, MC or SC, by choosing habitat size $L$ and introduction rate $\beta$ appropriately. Given $(\alpha, \mu)$, these parameter selections also determine the population-dynamic timescale $\langle\tau\rangle$. In simulation we varied the exogenous timescale, the environmental half-period $t_{1/2}$, from near-zero to levels far exceeding $\langle\tau\rangle$.

### Metrics

For parameter values we used, summed global densities $\rho_{tot}(t) = \rho_1(t) + \rho_2(t)$ exhibit little temporal variation. However, the difference of the species' densities, $m(t) = \rho_1(t) - \rho_2(t)$, proves relevant and informative (see Ripa and Ives, 2003). $m(t)$, with range $m(t) \in [-1, 1]$, indicates numerical superiority at time $t$. If $m(t) > 0$ persistently, species 1 dominates numerically; if $m(t) < 0$ persistently, species 2 dominates. If $m(t) \approx 0$, or fluctuates about 0, the competitors co-occur with roughly equal global densities. Figure 1 (A-D) shows two time series from our simulations, to illustrate the role of the density metrics. Note that $m(t)$ plays a role here analogous to the order parameter in ferromagnetic systems (Lo and Pelcovits, 1990; Tomé and de Oliveira, 1990; Chakrabarti and Acharyya, 1999).

Since we are interested in the competitive system's behavior in a periodically changing environment, a natural quantity to monitor is the period-averaged density difference (Tomé and de Oliveira, 1990; Sides *et al*., 1998a, b; Korniss *et al*.,2002; Buendia and Rikvold, 2008):

$$Q = \frac{1}{2t_{1/2}} \int_a^b m(t)\, dt \ ,$$ (1)

where an environmental period begins at $t = a$ and ends at $b = a + 2t_{1/2}$. The period-averaged density difference $Q$ is also referred to as the dynamic order parameter (Sides *et al*., 1998a; Korniss *et al*., 2000). For $t_{1/2} \gg \tau$, the environment changes slowly; decay of the respective resident's density is complete, and the system approaches competitive exclusion in each half period. Consequently, $m(t)$ reaches a limit cycle *symmetric* about zero, and in turn $Q \approx 0$. On the other hand, for $t_{1/2} \leq \tau$ the two species' densities do not have enough time to "switch" during each environmental half-period. Therefore, $m(t)$ reaches an asymmetric limit cycle, and in turn, $|Q| > 0$. Although, as we shall see in the next subsection, the scenarios can be more subtle than these two naïve cases, the quantity $Q$ provides an important measure of difference in abundance in periodically varying environments. In the simulations, we evaluated $Q$ by averaging $m(t)$ through each full period, and we define $\langle Q \rangle$ as the average of $Q$ over many successive periods. We also calculated the average absolute value, $\langle |Q| \rangle$, from the same data. Finally, we estimate a scaled, among-period variance of $Q$ (Sides *et al*., 1999; Korniss *et al*., 2000; Buendía and Rikvold, 2008):

$$X_L^Q = L^2 \left( \left\langle Q^2 \right\rangle_L - \left\langle |Q| \right\rangle_L^2 \right)$$ (2)

for different habitat sizes *L*.

These metrics do more than describe patterns in the $\rho_i(t)$. They identify transitions ("crossovers") in the competitive dynamics as functions of $t_{1/2}$. That is, $\langle |Q| \rangle$ declines rapidly toward zero, and $X_L^Q$ peaks, at critical values of $t_{1/2}$ where transitions in the dynamics occur (Sides *et al*., 1998a; Korniss *et al*., 2002; Robb *et al*., 2007; Buendía and Rikvold, 2008).

We also estimated the logarithmic growth rate of species *i* (*i* = 1, 2) as:

$$G[\rho_i(t)] = \ln L^2\, \rho_i(t+1) - \ln L^2\, \rho_i(t) = \ln \left[ \rho_i(t+1) \Big/ \rho_i(t) \right].$$ (3)

We were particularly interested in growth rates at low density, where clusters of the rare species should be small.

**Multi-cluster invasion in a periodic environment**

To begin, we let competitive advantage alternate periodically with $\beta = 10^{-4}$, and *L* = 90, 128, 180, and 256. In a *constant* environment these combinations of introduction rate and habitat size assure MC invasion with a lifetime $\langle \tau \rangle \approx 1240$. Note that for MC invasion, the endogenous timescale $\langle \tau \rangle$ is essentially independent of habitat size (Korniss and Caraco, 2005). We advance simulation time as Monte Carlo system steps. That is, during a single time unit we randomly select $L^2$ sites for updating *via* the individual-based model. Then each site, on average, updates once per time step.

Figure 2A plots $\langle |Q| \rangle$ as a function of $t_{1/2}$. $\langle |Q| \rangle$ declines toward zero for both very short ($t_{1/2} \ll \langle \tau \rangle$) and very long ($t_{1/2} \gg \langle \tau \rangle$) periods. At intermediate periodicity, when $t_{1/2}$ is "less than but within the same order of magnitude" as $\langle \tau \rangle$, which we write as $t_{1/2} \overset{\sim}{<} \langle \tau \rangle$, $\langle |Q| \rangle$ takes distinctly non-zero values. Hence the competitors' dynamics should exhibit two major transitions as $t_{1/2}$ increases - implying three different dynamical behaviors. Figure 2B verifies the transitions. The scaled variance $X_L^Q$ has two peaks (more pronounced as habitat size increases), each indicating a crossover from one dynamics to another.

Consider $t_{1/2} \ll \langle \tau \rangle$ first. The rapidly oscillating environment interrupts the dynamics' relaxation toward equilibrium each half-period; see Buceta *et al.* (2004). Hence, once both species have entered the habitat, neither can exclude the other (in ecological time), and we observe non-equilibrium coexistence. Figure 3A shows that the difference in global densities responds regularly, but with small magnitude, to each brief half-period's competitive asymmetry. Among periods the dynamics behaves, in effect, as if the time-varying environment had been replaced by its average (in physicists' terminology, the rapid, periodic environmental variation "anneals" propagation rates); see inset of Fig. 3A. In turn, the period-averaged density difference (or dynamic order parameter) $Q$ randomly fluctuates about zero (Fig. 3D). More carefully, reproductive advantage alternates so rapidly that the competitive interaction approaches the indeterminate case where the two species have identical demographic parameters. Over the long term, the species slowly exchange roles of more and less abundant ($m(t)$ changes sign, inset of Fig. 3A; see Fig. 10A), but each competitor's global density exceeds the background level maintained by introduction.

The non-equilibrium coexistence induced by rapid environmental fluctuation inhibits large-scale spatial order. Species 2 is rare initially. Introduced individuals of species 2 generate relatively small, randomly located clusters. These clusters ordinarily expand and contract a bit during consecutive half-periods, and some (by chance) grow large enough to persist lengthily. That is, they grow sufficiently large to maintain their single-species equilibrium inside, and localize between-species competition to their periphery (Allstadt *et al.*, 2007; 2009), promoting cluster longevity. Consequently, neither competitor produces the single large-scale cluster that precedes exclusion of the other species (Gandhi *et al.*, 1998; 1999; Yasi *et al.*, 2006). Figure 4, A and B, shows detail of the spatial system at the completion of consecutive half-periods. During the first half-period, species 2 had the greater propagation rate and advanced (Fig. 4A), and species 1 had the advantage during the second half-period (Fig. 4B). Local correlations are evident, but the system does not exhibit ordering at extended distances; coexistence results.

Secondly, we increase the half-period to match invasion time, so that $t_{1/2} \stackrel{\approx}{<} \langle \tau \rangle$ for MC parameters. Compared to the rapid environmental fluctuation just described, the reduced frequency of competitive reversal allows the dynamics to respond more strongly to the exogenous signal each half-period. Within most (but not all) intervals of length $2t_{1/2}$, the density-difference cycles from $m \approx 0.9$ to $m \approx 0$, and then returns to $m \approx 0.9$ (Fig. 3B). That is, species 2 advances from rarity during the half-period when it had the greater individual–level propagation rate, so that at the end of the half-period global densities are roughly equal (Fig. 5A). During the next half-period, the now superior species 1 grows and excludes species 2 (Fig. 5B). We refer to this pattern as an invasion-exclusion cycle. Since $\langle \tau \rangle$ is defined as the mean time elapsing until an inferior, resident species' density is halved, the system's behavior when $t_{1/2}$ and $\langle \tau \rangle$ are approximately equal follows from our understanding of invasion in a constant environment. The resident species maintains a greater density over the course of an entire environmental period, although it has no greater average propagation rate. Indeed, for thousands of consecutive periods, one species advances during one "season," only to be excluded the next – despite the two species' identical period-averaged demographic rates.

However, rough equivalence of the competitors' average densities appears over extended timescales. Although the same invasion-exclusion cycle repeats lengthily, noise can generate spontaneous density fluctuations large enough to "flip" global densities, so that the rare species becomes the more abundant. Figure 3E shows that $Q$, at apparently random times, switches from $Q \approx 0.6$ to $Q \approx -0.6$, reversing roles of common and rare species in less than $10^2$ periods. This lengthy-timescale switching between the two phases, each of which has a single, numerically dominant species throughout, is the manifestation of spontaneous symmetry-breaking in an interacting particle system (Binder and Heerman, 1997). This process, often referred to as a dynamic phase transition, has been well studied in physical systems with local interactions (*e.g.*, Korniss *et al*., 2000; Machado *et al*., 2005; Robb et al., 2007). Applying an insight from these studies yields an interesting ecological prediction; as a function of habitat size, the switching time between the two symmetric phases should increase faster than any power law. That is, the waiting time for fluctuations spanning the habitat, and consequently capable of switching the species' abundances, should increase exponentially with the size of the habitat (Goldenfeld, 1992). Figure 6 demonstrates this effect in our model. Matching of the fundamental timescales gives rise to symmetry breaking, *i.e.*, long-term dominance of one of the species (in very large habitats) despite the same mean demographic rates. At any given time, it is highly probable that one species holds numerical superiority; the other is introduced and then excluded each period. But we cannot predict, *a priori*, which species will be common.

Third, when the half-period exceeds the timescale of invasion, $t_{1/2} >> \langle \tau \rangle$, the currently superior species excludes the other following each environmental change. Invader clusters grow, coalesce, and the invader's global density advances to single-species equilibrium before roles reverse, *i.e.*, competitive exclusion is completed each half period. The symmetric limit cycle of $m(t)$ and associated $Q$ values are plotted in Figs. 3C and 3F, respectively.

**Growth rates and global densities**
Our model's preemptive competitors are equally subject to self-regulation and interspecific competition; hence we observe near constancy of summed densities during simulation (Fig. 1, C and D). Given the symmetry in the temporally varying propagation rates, each growth rate $G(\rho_i)$, averaged over time, should be zero. But the difference between the dynamics of non-equilibrium coexistence and invasion-exclusion cycles (*i.e.*, the absence/presence of time for the system to relax to equilibrium) might be revealed in growth-rate variation. Large samples of per-capitum growth rates reflect this difference only subtly. Figure 7 (A - D) plots $10^3$ logarithmic growth rates ($G[\rho_i(t)]$) of the initial resident (species 1) and initial invader (species 2), for both $t_{1/2} << \langle \tau \rangle$ and $t_{1/2} \stackrel{\approx}{<} \langle \tau \rangle$. We sampled each species regularly at intervals large enough to assure statistical independence.

Figures 7A and 7B iterate the point that under rapid environmental oscillation, nonequilibrium coexistence allows each species, over lengthy timescales, to explore the same range of global densities. Comparing Figs. 7A and 7C shows that for species 1 (the initial resident) neither the spatially averaged growth rates nor their levels of variability depend on that species' global density. For matching timescales (Figs. 7C and 7D) we observed the invader's growth rate more often at low density, a result of its initial rarity. Figure 7D indicates invasion-exclusion cycles of invader growth; growth-rate variability is larger for lower levels of $\rho_2(t)$. This is likely more than an effect of increased samples at low density. Invasion-exclusion cycles imply negative growth rates even at low density when species 2 has the lesser propagation rate, and faster growth when rare during half-periods with the greater propagation rate.

**Single-cluster invasion in a periodic environment**

Setting $\beta = 10^{-6}$ and letting $L = 64$ or $128$ assure SC invasion in a constant environment. Waiting times for successful introduction and invader nucleation are much longer for SC than for MC invasion (O'Malley *et al.*, 2005). For $L = 64$, we estimate invasion time $\langle \tau \rangle \approx 14000$, and estimate $\langle \tau \rangle \approx 6000$ for $L = 128$. Therefore, we simulated a greater range of $t_{1/2}$ for SC invasion than for the MC case.

Figure 8A plots $\langle |Q| \rangle$ against the environment's half-period. $\langle |Q| \rangle$ takes distinctly non-zero values at small $t_{1/2}$, and declines toward zero as period length increases. The scaled variance $X_L^Q$, Fig. 8B, has a single peak, for given habitat size $L$, where $t_{1/2} \approx \langle \tau \rangle$. The two results concur; the dynamics exhibits a single major transition as $t_{1/2}$ increases, crossing-over where the exogenous and endogenous timescales match.

For a rapidly alternating environment ($t_{1/2} << \langle \tau \rangle$), the rare species never can invade successfully (over $10^5$ to $10^6$ periods). Invader appearance and minimal spatial growth occur when the rare species has the greater propagation rate; see Figs.9A and 9D. But exclusion of the invader quickly follows after competitive advantage switches to the resident. The frequency of environmental change, combined with a very small introduction rate, inhibits formation of a critically-sized cluster of the rare species. Given an initially common species, the likelihood that the invader can establish a persistent cluster before losing competitive advantage and being excluded remains quite small.

When the environment changes very slowly, so that $t_{1/2} >> \langle \tau \rangle$, sufficient persistence of its propagation-rate advantage lets the currently superior competitor invade and exclude the other species in almost every half-period. Species' densities, hence $m(t)$, exhibit limit cycles, and $Q$ fluctuates about zero; see Figs. 9C and 9F.

For rapid environmental change, the invader cannot disrupt the resident's spatial order, even though the species are "on average" equivalent. For very slow environmental change, invasion and invader growth from rarity to numerical abundance are assured.

**Single-cluster invasion and stochastic resonance**

Finally, consider matching timescales ($t_{1/2} \approx \langle \tau \rangle$) in the SC regime. Figure 9B shows that $m(t)$ responds to environmental periodicity, but does not track it faithfully. In some half-periods the species with temporary competitive advantage advances from rarity and excludes the resident, but successful invasion in any given period is uncertain. That is, the dynamics unpredictably switches between exclusion of one species and exclusion of the other; see Fig. 9E. Hence the competitive system exhibits a transitional behavior between (essentially) no invasion ($t_{1/2} << \langle \tau \rangle$) and almost certain invasion each half-period ($t_{1/2} >> \langle \tau \rangle$).

Erratic switching between competitive-exclusion states when the two timescales match suggests stochastic resonance (Gammaitoni *et al.*, 1998). Consider a nonlinear system with two locally stable, equilibrium nodes, to which we add a relatively small periodic input (alternating propagation rates). Depending on the initial conditions, the dynamics will then oscillate gently around one or the other equilibrium; the system will not "switch" between alternative equilibria. Then we add noise; the system now has a non-zero switching probability. That is, with stochasticity added, the dynamics may move from one attractor to the other. This switching in the presence of noise, whether randomly erratic or approaching the pattern of the periodic input, defines stochastic resonance (Marchesoni *et al.*, 1996; Huppert and Stone, 1998; Sides *et al.*, 1998a; Vilar and Solé, 1998; Korniss *et al.*, 2002).

Our competition model does not exhibit bistability. In a constant environment, exclusion of the inferior species is the sole positive equilibrium. However, the competitively inferior resident can resist invasion for a long time, since competition is preemptive (Korniss and Caraco, 2005). The constant-environment model possesses a stable equilibrium and a metastable "quasi-equilibrium;" over a half-period, the latter may prevent the system from moving to the stable equilibrium. This sort of asymmetric system also can exhibit stochastic resonance when stability properties periodically reverse (Stocks *et al.*, 1993; Leung and Néda, 1999).

To demonstrate stochastic resonance in the simulation data, we estimated frequencies of waiting times elapsing between consecutive occurrences of equal species' abundance (*i.e.*, consecutive times when $m(t) = 0$). Designate the random waiting time $t_r$. For square-wave external variation, Korniss *et al.* (2002) assume that the time for growth of a nucleated cluster is negligible and approximate the probability density of $t_r$ analytically. The density, $f(t_r)$, has a shape symmetric about odd multiples of the half-period $t_{1/2}$:

$$f(t_r) = g(\Theta) \times \begin{cases} \sinh\left[(t_r/\langle\tau\rangle) - 2(n-1)\Theta\right] & for\ 2(n-1)t_{1/2} < t_r < (2n-1)t_{1/2} \\ \sinh\left[2n\Theta - (t_r/\langle\tau\rangle)\right] & for\ (2n-1)t_{1/2} < t_r < 2nt_{1/2} \end{cases} \quad (4)$$

where $\Theta = t_{1/2}/\langle\tau\rangle$, $g(\Theta) = \langle\tau\rangle^{-1} e^{-n\Theta}/(1 - e^{-\Theta})$, and $n = 1, 2, ...$ Figure 10 plots observed $f(t_r)$ for $t_{1/2} = 1.2 \times 10^4$. Observed frequencies peak at odd multiples of the half-period, the signature of stochastic resonance (Gammaitoni *et al.*, 1998; Acharyya, 1999; Sides *et al.*, 1998b; Korniss *et al.*, 2002), and match predicted $f(t_r)$ reasonably well.

## DISCUSSION

Our simulation study examines spatial competition in light of interaction between the timescale of environmental variation, relative to mean invasion time, and the spatial-clustering pattern underlying those invasions that succeed. When the environment varies slowly relative to the population-dynamic timescale, multi-cluster and single-cluster invasion processes predict the same qualitative behavior. Each half-period the currently advantaged competitor invades and excludes the resident; the environment never interrupts the endogenous relaxation to equilibrium. However, the timescale-difference interacts with invasion regime when the environmental half-period matches, or is less than, mean invasion time. We observe a range of dynamics, from exclusion of the rare species to unpredictable exclusion of the resident.

Figure 11A shows time-dependent densities for a rapidly alternating environment in the MC-regime. The two species are effectively equivalent competitively, and we observe non-equilibrium coexistence. Decreasing the pace of environmental change to match invasion time (Fig. 11B) generates invasion-exclusion cycles, since the longer half-period often permits relaxation to the currently favored single-species equilibrium. For the SC-regime, Fig. 11C shows the results for rapid environmental variation; the rare species cannot generate a cluster large enough to invade. Finally, increasing the half-period length to match invasion time in the SC-regime generates stochastic resonance; see Fig. 11D.

When the environmental and endogenous timescales match, the long-term dynamics is least predictable. In the MC-regime the short-term dynamics is dominated by invasion-exclusion cycles driven by introduction events and alternating competitive advantage. However, randomly (though rarely) the species quickly exchange roles of common and rare. The period-averaged density difference ($Q$) exhibits little short-term variation, but varies bimodally over the long term (Fig. 3E). Hence the norm of the dynamic order parameter $\langle|Q|\rangle$ (see subsection on Metrics) best captures the symmetry breaking associated with the dynamic phase transition from numerical dominance by one species to dominance by its competitor. Deterministic environmental periodicity and stochastic spatial propagation combine to render long-term prediction uncertain for MC-invasion. For the SC-regime, uncertainty inherent to stochastic resonance is summarized by the distribution of waiting rimes between switches in roles of common and rare species (Fig. 10).

We began simulations with one species common and the other at zero density, focusing on the dynamics of rarity. We wondered if the invader's consistent failure to advance in the SC-regime ($t_{1/2} << \langle \tau \rangle$) and/or the exclusion-invasion cycles in the MC-regime ($t_{1/2} \stackrel{\approx}{<} \langle \tau \rangle$) were consequences of initial conditions. We repeated these simulations (only once each) with initial condition $\left[ \rho_1 = 0.5, \rho_2 = 0.48 \right]$. The results, presented in Fig. 12, closely resemble those in Fig. 11, after transients disappear. We conclude that the dynamics does not depend on initial conditions.

Any approximation to our model that assumes strong spatial mixing will fail to produce the observed range of dynamics (Antonovics *et al.*, 2006). Our model's properties of discreteness and stochasticity define its fundamental character (O'Malley et al., 2009), and the dynamics of rarity is studied realistically by assuming discrete individuals and stochastic demographic events (Durrett and Levin, 1994; Ellner *et al.*, 1998; Duryea *et al.*, 1999; Escudero *et al.*, 2004).

We described our work as examining interaction of the difference between timescales with invasion regime. This categorization organized our simulation study. However, scaling arguments of nucleation theory allow us to simplify the description as interaction between the exogenous timescale $t_{1/2}$ and the introduction rate $\beta$, since both the characteristic time scale $\langle \tau \rangle$ and the characteristic length scale $D$ of the endogenous dynamics depend on $\beta$.

Consider MC-invasion in a constant environment, and let habitat size $L$ grow large. Then $\langle \tau \rangle \sim \left( I \nu^2 \right)^{-1/3}$, where $I$ is the nucleation rate/unit area, and $\nu$ is the radial velocity at which nucleated clusters of the superior competitor grow. We noted above that $\langle \tau \rangle \sim \beta^{-1/3}$, since $I \sim \beta$ (see Appendix 1). That is, the characteristic time scale for MC-invasion increases as the inverse of the cube root of the introduction rate (Korniss and Caraco, 2005). The characteristic length scale $D$, the expected distance between nucleated invader clusters, scales as $D \sim \left( \nu / I \right)^{1/3}$, and we have $D \sim \beta^{-1/3}$ (O'Malley *et al.*, 2006a). That is, the characteristic length scale for MC-invasion also increases as the inverse of the cube root of the introduction rate.

Propagation and mortality rates are fixed in a constant environment. From above, for any fixed habitat size $L$, there is a critical introduction rate $\beta_C$ such that for $\beta < \beta_C$, we have $D \sim \beta^{-1/3} >> L$. Then constant-environment invasion crosses over to the SC-regime where $\langle \tau \rangle \sim \beta^{-1}$. Therefore, both the endogenous timescale and invasion mode (given habitat size) depend on the introduction rate, and our results reveal an interaction between $t_{1/2}$ and $\beta$.

**Model context**

Our model assumes rare, but repeating, introduction of individuals, an assumption prompted by observed invasions of non-native species (Veltman *et al.*, 1996; Loreau and Mouquet, 1999; Sax and Brown, 2000). Not every introduction succeeds; those that do initiate spatially clustered growth. The assumption of spatially explicit introduction contrasts with models treating a rare species' as spread uniformly across an environment (Snyder and Chesson, 2004). The period-averaged competitive symmetry of our model's two species is less realistic. However, the results offer a comparison for spatial competition where a superior species' advantage varies temporally.

A number of previous models examine competition under temporal variation; this study and Chan *et al*. (2009) differ by taking an individual-based perspective. Chesson's (1990) study represents a standard class of models for competition under temporal variation. Seed-bank densities of two, homogeneously mixing plant species are tracked in discrete time. A varying environment and density-dependence affect each species' annual growth rate/unit density. That is, the fraction of a species' seeds germinating varies randomly between generations, and competition reduces the seed yield/adult. Analysis focuses on growth-rate responses to competition in good *vs* bad environments, and on differences in the growth rates' response to increasing competition across environments. This type of model permits greater between-species differences than do our assumptions; coexistence mechanisms are therefore more general. Model construction assumes a single scale for temporal variation; the environment changes randomly, independently each discrete generation. However, autocorrelation can be used to manipulate the exogenous time scale (Chesson, 1990).

Another approach considers timescale interaction explicitly, but retains the homogeneous-mixing assumption. Abrams (2004) develops a continuous-time model where two consumer species compete for a common resource that has its own dynamics with a periodic growth rate. Trade-offs can promote persistence. If the species with lower efficiency at high resource density has the greater efficiency at low resource density, coexistence can result. Abrams (2004) focuses on the rapidity of the consumers' demographic responses to temporal variation in resource density, and concludes that interaction of endogenous population dynamics and periodic environmental variation impacts patterns in competitor abundances. Recer *et al*. (1987) and Cross *et al*. (2005) make similar points for simpler systems.

Schoolmaster and Snyder (2007; see Snyder, 2007) combine temporal and spatial variation in a model for competing perennial plants occupying a 1-dimensional environment. An adult's seed dispersal and competitive suppression of seedling establishment both decline with distance from the adult. Establishment further depends on an environmental quality that varies periodically both in time and along the linear environment. Each periodicity has its own scale. The authors ask how these fluctuations, compared to environmental constancy, affect the growth rate of a rare species dispersed at uniformly along the environment. The results suggest that the impact of environmental variation depends on its interaction with the competitors' life history traits (Schoolmaster and Snyder, 2007), hence interaction with the population dynamic timescale. A series of different models addressing interspecific competition agree that quantitative and qualitative predictions can depend on the interaction of endogenous and exogenous timescales.

Chan *et al*. (2009) model individual-based competition between species. Each species' birth rate varies seasonally, the species' respective death rates are fixed, and introduction is not considered explicitly. The authors seek analytical conditions for coexistence. To do so, they relax spatial structuring of offspring dispersal, a key assumption of our study. Given a sufficiently large dispersal neighborhood, Chan *et al*. (2009) prove that temporal variation may promote coexistence, which is not found absent the variation. Sufficient conditions for coexistence match those of the homogeneous mean-field approximation to their model.

**Cross-disciplinary integration**

Our results demonstrate that the impact of a periodic environment on spatial competition can depend on the difference between exogenous and endogenous timescales, and that the resulting dynamics can be complex. Similar analyses of interacting timescales have successfully advanced understanding of spatially structured physical systems (Chakrabarti and Acharyya, 1999). Particular examples include time-dependent properties of ferromagnetic thin films in an oscillating external magnetic field (Korniss *et al*., 2000; Robb *et al*., 2007, 2008; Buendia and Rikvold, 2008), and behavior of a catalytic reaction subject to periodic variation in CO pressure (Machado *et al*., 2005). These systems can exhibit the symmetry breaking we found when the endogenous and exogenous timescales matched in the MC-invasion regime. Details of local interactions differ substantially among these physical and spatial models in ecology. But they share important dynamical behaviors emerging from timescale interaction. We hope that the convergent predictions suggest a more general understanding.

## ACKNOWLEDGEMENTS

This material is based upon work supported by the National Science Foundation under Grants DEB-0918413 (GK) and DEB-0918392 (TC).

**REFERENCES**

Abrams, P.A. 2004. When does periodic variation in resource growth allow robust coexistence of competing consumer species? *Ecology* 85:372-382.

Acharyya, M. 1999. Nonequilibrium phase transition in the kinetic Ising model: Existence of a tricritical point and stochastic resonance. *Phys. Rev. E* 59:218—221.

Adler, P.B. and Drake, J.M. 2008. Environmental variation, stochastic extinction, and competitive coexistence. *Am. Nat*. 172:E186-E195.

Allstadt, A., Caraco, T. and Korniss, G. 2007. Ecological invasion: spatial clustering and the critical radius. *Evol. Ecol. Res*. 9:375-394.

Allstadt, A., Caraco, T. and Korniss, G. 2009. Preemptive spatial competition under a reproduction-mortality constraint. *J. Theor. Biol*. 258:537-549.

Altizer, S., Dobson, A., Hosseini, P., Hudson, P., Pascual, M. and Rohani, P. 2006. Seasonality and the dynamics of infectious diseases. *Ecol. Lett*. 9:467-484.

Antonovics, J., McKane, A.J. and Newman, T.J. 2006. Spatiotemporal dynamics in marginal populations. *Am. Nat*. 167:16-27.

Avrami, M. 1940. Kinetics of phase change. II. Transformation-time relations for random distribution of nuclei. *J. Chem. Phys*. 8:212-240.

Avrami, M. 1941. Geometry and dynamics of populations. *Philos. Science* 8:115-132.

Binder, K. and Heerman, D.W. 1997. *Monte Carlo Simulations in Statistical Physics. An Introduction*. 3$^{rd}$ Ed. Berlin: Springer,

Bolker, B. and Pacala, S.W.. 1999. Spatial moment equations for plant competition: understanding spatial strategies and the advantage of short dispersal. *Am. Nat*.153:575-602.

Buceta, J., Escudero, C., de la Rubia, F.J. and Lindenberg, K. 2004. Outbreaks of Hantavirus induced by seasonality. *Phys. Rev. E* 69:021906, 8 pp.

Buendía, G.M. and Rikvold, P.A. 2008. Dynamic phase transition in the two-dimensional Ising model in an oscillating field: universality with respect to the stochastic dynamics. *Phys. Rev. E* 78:051108, 7 pp.

Caraco, T., Glavanakov, S., Li, S., Maniatty, W. and Szymanski, B.W. 2006. Spatially structured superinfection and the evolution of disease virulence. *Theor. Pop. Biol*. 69:367-384.

Caswell, H. and Cohen, J.E. 1995. Red, white and blue: environmental variance spectra and coexistence in metapopulations. *J. Theor. Biol*. 176: 301-316.

Chakrabarti, B.K. and Acharyya, M. 1999. Dynamic transitions and hysteresis. *Rev. Mod. Phys.* 71:847—859.

Chan, B., Durrett, R. and Lanchier, N. 2009. Coexistence for a multitype contact process with seasons. *Ann. Appl. Prob*. 19:1921-1943.

Chesson, P. 1990. Geometry, heterogeneity, and competition in variable environments. *Phil. Trans. R . Soc. London* B330:175-194.

Chesson, P. 2000. General theory of competitive coexistence in spatially-varying environments. *Theor. Pop. Biol*. 58:211-237.

Chesson, P. and Huntly, N. 1997. The roles of harsh and fluctuating conditions in the dynamics of ecological communities. *Am. Nat*. 150:165-173.

Connolly, S. R. and Muko, S. 2003. Space preemption, size-dependent competition, and the coexistence of clonal growth forms. *Ecology* 84:2979-2988.

Cross, P.C., Lloyd-Smith, J.O., Johnson, P.L.F. and Getz, W.M. 2005. Duelling timescales of host movement and disease recovery determine invasion of disease in structured populations. Ecol. Lett. 8:587-595.

D'Antonio, C. M. 1993. Mechanisms controlling invasion of coastal plant communities by the alien succulent *Carpobrotus edulis*. *Ecology* 74:83-95.

Descamps-Julien, B. and Gonzalez, A. 2005. Stable coexistence in a fluctuating environment: an experimental demonstration. *Ecology* 86:2815-2824.

Dieckmann, U., Law, R. and Metz, J.A.J., eds. 2000. *The Geometry of Ecological Interactions*: *Simplifying Spatial Complexity*. Cambridge, UK: Cambridge University Press.

D'Odorico, P., Laio, F., Ridolfi, L. and Lerdau, M.T. 2008. Biodiversity enhancement induced by environmental noise. *J. Theor. Biol*. 255:332-337.

Durrett, R. and Levin, S.A. 1994. The importance of being discrete (and spatial). *Theor. Pop. Biol*. 46:363–394.

Duryea, M., Caraco, T., Gardner, G., Maniatty, W. and Szymanski, B.K. 1999. Population dispersion and equilibrium infection frequency in a spatial epidemic. *Physica D* 132: 511–519.

Ellner S.P., Sasaki A., Haraguchi Y. and Matsuda, H. 1998. Speed of invasion in lattice population models: pair-edge approximation. *J. Math. Biol*. 36: 469-484.

Escudero, C., Buceta, J., de la Rubia, F.J. and K. Lindenberg, K. 2004. Extinction in population dynamics. *Phys. Rev.E* 69:021908, 9pp.

Gammaitoni, L., Marchesoni, F. and Santucci, S. 1995. Stochastic resonance as a bona fide resonance. *Phys. Rev. Lett*.74:1052, 4 pp.

Gandhi, A., Levin, S. and Orszag, S. 1998. "Critical slowing down" in time-to-extinction: an example of critical phenomena in ecology. *J. Theor. Biol*. 192:363-376.

Gandhi, A., Levin, S. and Orszag, S. 1999. Nucleation and relaxation from meta-stability in spatial ecological models. *J. Theor. Biol*. 200:121-146.

Goldenfeld, N. 1992. *Lecture Notes on Phase Transitions and the Renormalization Group*. Addison-Wesley, New York, NY.

Greenman, J.V. and Norman, R.A. 2007. Environmental forcing, invasion and control of ecological and epidemiological systems. *J. Theor. Biol*. 247:492-506.

Herben, T., During, H.J. and Law, R. 2000. Spatio-temporal patterns in grassland communities. Pp. 48-64 in Dieckmann, U., Law, R. and Metz, J.A.J., eds. *The Geometry of Ecological Interactions: Simplifying Spatial Complexity*. IIASA: Vienna, AU.

Herrick, J., Jun, S. Bechhhoefer J. and Bensimon, A. 2002. Kinetic model of DNA replication in eukaryotic organisms. *J. Mol. Biol*. 320:741-750.

Holt, R. D. and Barfield, M. 2003. Impacts of temporal variation on apparent competition and coexistence in open ecosystems. *Oikos* 101:49--58.

Holt, R.D., Barfield, M. and Gomulkiewicz, R. 2004. Temporal variation can facilitate niche evolution in harsh sink environments. *Am. Nat*. 164:187-200.

Holway, D. A. 1998. Factors governing rate of invasion: a natural experiment using Argentine ants. *Oecologia* 15:206-212.

Huppert, A., and Stone, L. 1998. Chaos in the Pacific's coral reef bleaching cycle. *Am. Nat*. 152:447-459.

Hurt, G.C. and Pacala, S.W. 1995. The consequences of recruitment limitation: reconciling chance, history, and competitive differences between plants. *J. Theor. Biol*. 176:1-12.

Hutchinson, G. E. 1961. The paradox of the plankton. *Am. Nat*. 93:145-159.

Ives, A.R. 1995. Predicting the response of populations to environmental change. *Ecology* 76:1039-1052.

Johnson W. A. and Mehl, R.F. 1939. Reaction kinetics in processes of nucleation and growth. *Trans. Am. Inst. Mining Metallurgic Engineer*. 135:416-442.

Jun, S., Herrick, J., Bensimon, A. and Bechhoefer, J. 2004. Persistence length of chromatin determines origin spacing in Xenopus early-embryo DNA replication. *Cell Cycle* 3:223-229.

Kaschiev, D. 2000. *Nucleation: Basic Theory with Applications*. Oxford: Butterworth-Heinemann.

Kolmogorov, A. N. 1937. A statistical theory for the recrystallization of metals. *Bull. Acad. Sci*., *USSR, Physics Ser*. 1:355-359.

Korniss, G., White, C.J., Rikvold, P.A. and Novotny, M.A. 2000. Dynamic phase transition, universality, and finite-size scaling in the two dimensional kinetic Ising model in an oscillating field. *Phys. Rev. E* 63:016120, 5 pp.

Korniss, G., Rikvold, P.A. and Novotny, M.A. 2002. Absence of first-order transitions and tri-critical point in the dynamic phase diagram of a spatially extended bistable system in an oscillating field. *Phys. Rev. E* 66:056127, 5 pp.

Korniss, G. and Caraco, T. 2005. Spatial dynamics of invasion: the geometry of introduced species. *J. Theor. Biol.* 233:137-150.

Lehman, C.L. and Tilman, D. 1997. Competition in spatial habitats. Pp. 185-203 in: Tilman, D. and Kareiva, P., eds. *Spatial Ecology: The Role of Space in Population Dynamics and Interspecific Interactions*. Princeton: Princeton University Press.

Leung, K.-T. and Néda, Z. 1999. Nontrivial stochastic resonance temperature for the kinetic Ising model. *Phys. Rev. E* 59:2730-2735.

Lo, W.S. and Pelcovits, R.A. 1990. Ising model in a time-dependent magnetic field. *Phys. Rev. A* 42:7471—7474.

Loreau, M. and de Mazancourt, C. 2008. Species synchrony and its drivers: neutral and nonneutral community dynamics in fluctuating environments. *Am. Nat.* 172:E48-E66.

Loreau, M. and Mouquet, N. 1999. Immigration and the maintenance of local species diversity. *Am. Nat.* 154:427-440.

Machado, E., Buendia, G.M., Rikvold, P.A. and Ziff. R.M. 2005. Response of a catalytic reaction to periodic variation of the CO pressure: Increased $CO_2$ production and dynamic phase transition. *Phys. Rev. E* 71:016120, 7pp.

Mágori, K., Szabó, P., Mizera, F. and Meszéna, G. 2005. Adaptive dynamics on a lattice: role of spatiality in competition, co-existence and evolutionary branching. *Evol. Ecol. Res.* 7: 1-21.

Marchesoni, F., Gammaitoni, L. and Bulsara, A.R. 1996. Spatiotemporal stochastic resonance in a $\phi^4$ model of kink-antikink nucleation. *Phys. Rev. Lett.* 76:2609, 4 pp.

Moro, E. 2001. Internal fluctuations' effects on Fisher waves. *Phys. Rev. Lett.* 87:238303, 4 pp.

Neubert, M.G., Kot, M. and Lewis, M.A. 2000. Invasion speeds in fluctuating environments. *Proc. R. Soc. London* B 267:1603-1610.

O'Malley, L., Allstadt, A., Korniss, G. and Caraco, T. 2005. Nucleation and global time scales in ecological invasion under preemptive competition. Pp. 117-124 in Stocks, N., Abbott, G.D. and Morse, R.P., eds. *Fluctuations and Noise in Biological, Biophysical, and Biomedical Systems III*. Proceedings of SPIE, Vol. 5841, Bellingham, WA: SPIE.

O'Malley, L., Basham, J., Yasi, J.A., Korniss, G., Allstadt, A. and Caraco, T. 2006a. Invasive advance of an advantageous mutation: nucleation theory. *Theor. Pop. Biol.* 70:464-478.

O'Malley, L., Kozma, B., Korniss, G., Rácz, Z. and Caraco, T. 2006b. Fisher waves and front propagation in a two-species invasion model with preemptive competition. *Phys. Rev. E* 74:041116, 7 pp.

O'Malley, L., Korniss, G. and Caraco, T. 2009. Ecological invasion, roughened fronts, and a competitor's extreme advance: integrating stochastic spatial-growth models. *Bull. Math. Biol.* 71:1160-1188.

Platt, W.J. and Weis, I.M., 1985. An experimental study of competition among fugitive prairie plants. *Ecology* 66:708-720.

Rácz, E.V.P. and Karsai, J. 2006. The effect of initial pattern on competitive exclusion. *Comm. Ecol.*7:23-33.

Ramos, R. A., Rikvold, P.A. and Novotny, M.A. 1999. Test of the Kolmogorov-Johnson-Mehl-Avrami picture of meta-stable decay in a model with microscopic dynamics. *Phys. Rev .B* 59:9053-9069.

Recer, G.M., Blanckenhorn, W.U., Newman, J.A., Tuttle, E.M., Withiam, M.L. and Caraco, T. 1987. Temporal resource variability and the habitat-matching rule. *Evol. Ecol*. 1:363-378.

Rikvold, P. A., Tomita, H., Miyashita, S. and Sides, S.W. 1994. Metastable lifetimes in a kinetic Ising model: dependence on field and system size. *Phys.Rev. E* 49:5080--5090.

Ripa, J. and Ives, A.R. 2003. Food web dynamics in correlated and autocorrelated environments. *Theor. Pop. Biol*. 64:369-384.

Robb, D.T., Rikvold, P.A., Berger, A. and Novotny, M. 2007. Conjugate field and fluctuation-dissipation for the dynamic phase transition in the two-dimensional kinetic Ising model. *Phys. Rev. E* 76:021124, 10 pp.

Robb, D.T, Xu, Y.H., Hellwig, O., McCord, J., Berger, A., Novotny, M.A. and Rikvold, P.A. 2008. Evidence for a dynamics phase transition in [Co/Pt]$_3$ magnetic monolayers. *Phys. Rev. B* 78:134422, 11pp.

Sax, D. F. and Brown, J. H., 2000. The paradox of invasion. *Global Ecol. Biogeo*. 9, 361-371.

Schoolmaster, D.R. and Snyder, R.E. 2007. Invasibility in a spatiotemporally fluctuating environment is determined by the periodicity of fluctuations and resident turnover rates. *Proc. R. Soc. B* 274:1429-1435.

Sides, S. W., Rikvold, P.A. and Novotny, M.A. 1998a. Kinetic Ising model in an oscillating field: finite-size scaling at the dynamic phase transition. *Phys. Rev. Lett.* 81:834-837.

Sides, S. W., Rikvold, P.A. and Novotny, M.A. 1998b. Stochastic hysteresis and resonance in a kinetic Ising system. *Phys. Rev. E* 57:6512—6533.

Sides, S. W., Rikvold, P.A. and Novotny, M.A. 1999. Kinetic Ising model in an oscillating field: Avrami theory for the hysteretic response and finite-size scaling for the dynamic phase transition. *Phys. Rev. E* 59:2710—2729.

Snyder, R.E. 2007. Spatiotemporal population distributions and their implications for species coexistence in a variable environment. *Theor. Pop. Biol*. 72:7-20.

Snyder, R.E. and Chesson, P. 2004. How the spatial scales of dispersal, competition, and environmental heterogeneity interact to affect coexistence. *Am. Nat.* 164:633-650.

Spencer, C.C., Saxer, G., Travisano, M. and Doebeli, M. 2007. Seasonal resource oscillations maintain diversity in bacterial microcosms. *Evol. Ecol. Res*. 9:775-787.

Stocks, N. D., Stein, N.D. and McClintock, P.V.E. 1993. Stochastic resonance in monostable systems. *J. Physics A* 26:L385-L390.

Tilman, D., and Kareiva, P., eds. 1997. *Spatial Ecology: The Role of Space in Population Dynamics and Interspecific Interactions*. Princeton: Princeton University Press.

Tomé, T. and de Oliveira, M.J. 1990, Dynamic phase transition in the kinetic Ising model under a time-dependent oscillating field. *Phys. Rev. A* 41:4251—4254.

Travis, J.M.J., Hammershøj, M. and Stephenson, C. 2005. Adaptation and propagule pressure determine invasion dynamics: insights from a spatially explicit model for sexually reproducing species. *Evol. Ecol. Res*. 7:37-51.

van Saarloos, W. 2003. Front propagation into unstable states. *Phys. Rep*. 386:29-222.

Veltman, C. J., Nee, S. and Crawley, M. J., 1996. Correlates of introduction success in exotic New Zealand birds. *Am. Nat.* 147, 542-557.

Vilar, J. M. G. and R. V. Solé. 1998. Effects of noise in symmetric two-species competition. *Phys. Rev. Lett.* 80:4099-4102.

Wilson, W.G. 1996. Lotka's game in predator-prey theory: linking populations to individuals. *Theor. Pop. Biol*. 50:368-393.

Wilson, W. 1998. Resolving discrepancies between deterministic population models and individual-based simulations. *Am. Nat*. 151:116-134.

Yasi, J. A., Korniss, G. and Caraco, T. 2006. Invasive allele spread under preemptive competition. Pp. 165-169 in *Computer Simulation Studies in Condensed Matter Physics*

*XVIII*. Landau, D.P., Lewis S.P. and H.-B. Schüttler, eds. Springer Proceedings in Physics Vol. 105, Heidelberg: Springer.

Yurkonis, K.A. and Meiners, S.J. 2004. Invasion impacts local species turnover in a successional system. *Ecol. Lett*. 4:764-769.

Zhang, H. and Bechhoefer, J. 2006. Reconstructing DNA replication kinetics from small DNA fragments. *Phys. Rev. E* 73:051903, 9 pp.

## APPENDIX 1

Summarizing model transitions at an arbitrary site $\mathbf{x}$ in a constant environment, we have:

$$0 \overset{\beta + \alpha_1 \eta_1(\mathbf{x})}{\rightarrow} 1, \quad 0 \overset{\beta + \alpha_2 \eta_2(\mathbf{x})}{\rightarrow} 2, \quad 1 \overset{\mu}{\rightarrow} 0, \quad 2 \overset{\mu}{\rightarrow} 0, \tag{A.1}$$

where 0, 1, 2 indicates whether the site is empty, resident-occupied, or invader-occupied, respectively. In a constant environment, simulations always reveal strongly clustered growth of the competitively superior invader. Once an invader cluster reaches the critical radius, that cluster, on average, grows approximately deterministically with radial velocity $v$.

For small environments or for sufficiently small introduction rates, so that $L \ll D$, invasion almost always occurs through spread of a single invader cluster. That is, a habitat size much smaller than the mean distance between clusters (the mean we would observe in an infinite environment) implies the SC regime. In simulation we confirmed that nucleation of a successful invading cluster is a Poisson process with nucleation rate per unit area $I$. In the SC regime, the lifetime of the resident species $\tau$ is dominated by the lengthy waiting time until the first successful invader cluster nucleates. Differences between competitors' propagation and mortality rates govern the time required for the cluster to grow and exclude the resident, but this time is short compared to the waiting time for nucleation (O'Malley *et al*., 2006a). Therefore, if $t_n$ is the exponential waiting time until nucleation occurs (the first invader cluster reaching $R_{crit}$), then $\langle \tau \rangle \approx \langle t_n \rangle = \left( L^2 I \right)^{-1}$ in the SC regime. The $\beta$-dependence of the nucleation rate per unit area is $I \sim \beta$. Therefore, $\langle t_n \rangle \approx \left( L^2 \beta \right)^{-1}$; waiting time for invasion varies inversely with habitat area and the introduction rate.

Foe SC invasion, the cumulative distribution of invasion times $\Pr\left[ \tau > t \right]$, *i.e.*, the probability that the resident's global density has not decayed to $\rho_1^* / 2$ by time $t$ is a modification of the 0-term of a Poisson probability function :

$$P_{not}(t) = \begin{cases} 1 & \text{for } t \leq t_g \\ \exp[-(t - t_g)/\langle t_n \rangle] & \text{for } t > t_g \end{cases} \tag{A.2}$$

In the SC-regime, $t_g \sim L / v$ is the approximately deterministic growth time until the invading species drives the resident to half its initial density. For very small nucleation rates per unit area, $\langle t_n \rangle \gg t_g$. Therefore, the lifetime of the resident $\tau$ is governed by the large average waiting time until the first successful invader cluster nucleates, so that $\langle \tau \rangle = \langle t_n \rangle + t_g \approx \langle t_n \rangle = (L^2 I)^{-1}$ in the SC regime.

In large environments (or for large introduction rates), $\Pr\left[ \tau > t \right]$ approaches a step-function centered on the system size-independent lifetime $\langle \tau \rangle$. In the limit of a large environment, we can approximate global densities closely by Avrami's Law, or KJMA theory (Kolmogorov, 1937; Johnson and Mehle, 1939; Avrami, 1940):

$$\rho_1(t) = \rho_1^* \exp\left[ -\ln(2) \left( \frac{t}{\langle \tau \rangle} \right)^3 \right] \quad \text{and} \quad \rho_2(t) = \rho_2^* \left[ 1 - \exp\left[ -\ln(2) \left( \frac{t}{\langle \tau \rangle} \right)^3 \right] \right], \tag{A.3}$$

where $\langle \tau \rangle \propto (Iv^2)^{-1/3}$. An important result in nucleation theory, Avrami's law provides the generic functional form of the time-dependent global densities during MC invasion. Further, the parameters of a specific model for spatially structured ecological interactions (i.e., the local transition rates $\alpha_1$, $\alpha_2$, $\mu$, and $\beta$) govern the characteristic time scale (the lifetime $\langle \tau \rangle$) through their impact on the nucleation rate per unit area $I(\alpha_1, \alpha_2, \mu, \beta)$ and the invader-cluster radial velocity $v(\alpha_1, \alpha_2, \mu)$. Thus, Eq. (A.3) also identifies an important connection between model-specific processes at the level of individual propagation/mortality rates and pattern at the population-dynamic level. We previously linked these two organizational levels through the nucleation rate $I(\alpha_1, \alpha_2, \mu, \beta)$ and the invader-cluster radial velocity $v(\alpha_1, \alpha_2, \mu)$; see O'Malley *et al.* (2006a, 2009).

Our paper examines a specific model, but the framework of nucleation theory has broad ecological significance. Any time-homogenous invasion processes combining rare introduction, preemptive competition, and localized propagule dispersal (*i.e.*, strong dispersal limitation) in a large environment will likely generate spatially clustered growth and global dynamics consistent with nucleation theory's predictions (O'Malley *et al.*, 2006a, 2009; Allstadt *et al.*, 2007). That is, in a constant environment nucleation theory should predict not only equilibrium states, but also time-dependent global densities, when locally dispersing species compete for space.

**FIGURES**

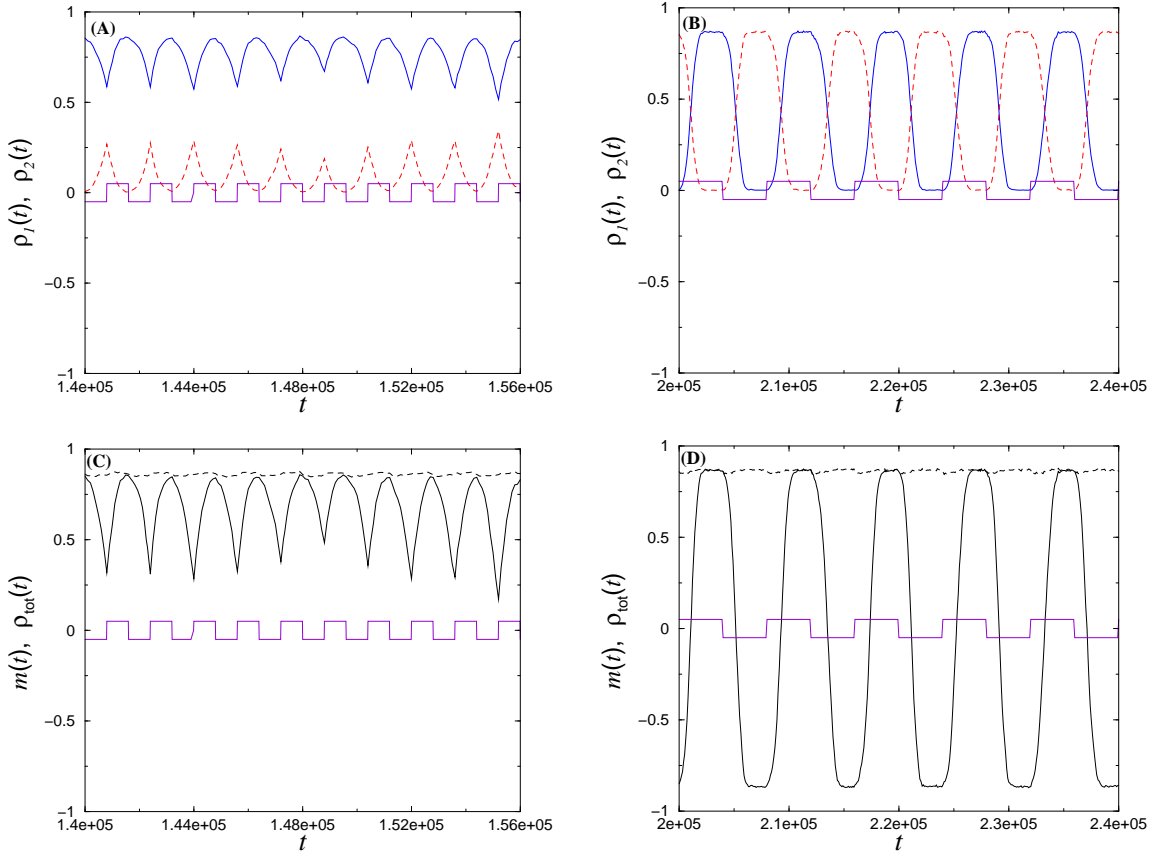

**Fig. 1**. Time-dependent densities $\rho_i(t)$ $(i = 1, 2)$, density difference $m(t) = \rho_1(t) - \rho_2(t)$, and total density $\rho_{\text{tot}}(t) = \rho_1(t) + \rho_2(t)$ in a periodically changing environment; the square wave illustrates $\Delta(t)$. $\beta = 10^{-4}$ and habitat size $L = 128$ in each plot. Here, and throughout this study, the mortality rate is $\mu = 0.1$. The invasion time for these parameter values is $\langle \tau \rangle \approx 1240$, and is essentially independent of the habitat size for sufficiently large ($L > 64$) habitats. (A) Time-dependent densities for $t_{1/2} = 800 \stackrel{\approx}{<} \langle \tau \rangle$; (B) for $t_{1/2} = 4000 >> \langle \tau \rangle$ (solid curve is $\rho_1(t)$ and dashed curve is $\rho_2(t)$). (C) Time-dependent density difference (solid curve) and total density (dashed curve) for $t_{1/2} = 800 \stackrel{\approx}{<} \langle \tau \rangle$; (D) for $t_{1/2} = 4000 >> \langle \tau \rangle$.

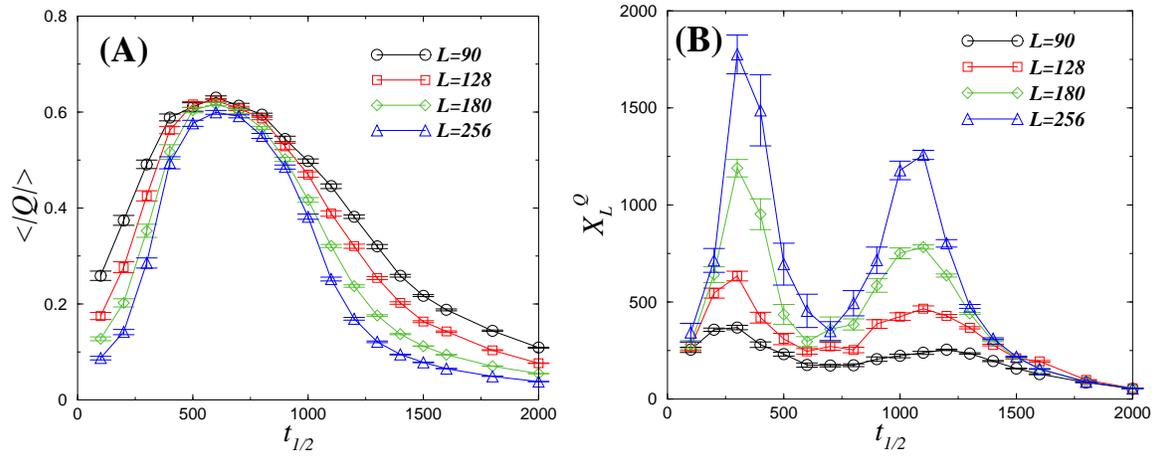

**Fig. 2**. Metrics for multi-cluster invasion. (A) Average of absolute value of $Q$ as function of environmental half-period $t_{1/2}$. $\beta = 10^{-4}$; habitat size $L = 90, 128, 180, 256$. $\langle \tau \rangle \approx 1240$ for each of these habitat sizes. (B) The scaled variance of $Q$. Peaks indicate values of $t_{1/2}$ where transitions in competitive dynamics occur.

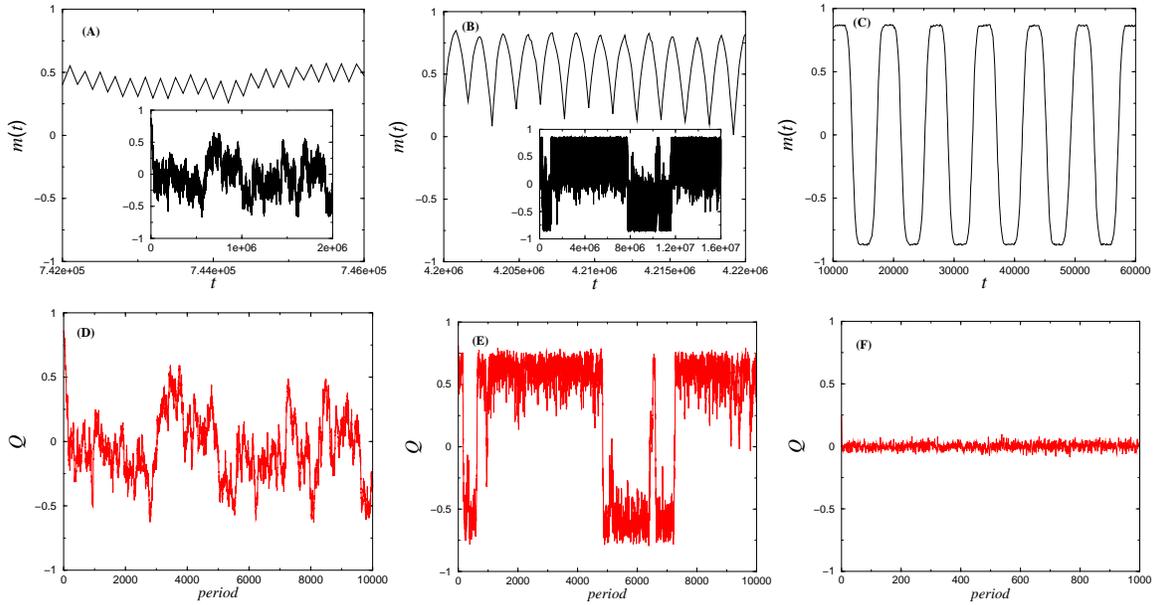

**Fig. 3**. Multi-cluster regime: segments of the density difference $m(t)$ as a function of time, and period-averaged density difference $Q$ plotted for a large number of consecutive periods. $\beta = 10^{-4}$ and habitat size $L = 128$. (A) $t_{1/2} = 100 \ll \langle \tau \rangle$. Note scale of ordinate. Inset: Long-term plot, indicating non-equilibrium coexistence with roughly equal averaged densities. (B) $t_{1/2} = 800 \overset{\approx}{<} \langle \tau \rangle$. Invasion-exclusion cycles. Inset: Long-term plot. (C) $t_{1/2} = 4000 \gg \langle \tau \rangle$. System reaches single-species equilibrium during each, lengthy half-period. (D) $t_{1/2} = 100 \ll \langle \tau \rangle$; each period lasts 200 time units. (E) $t_{1/2} = 800 \overset{\approx}{<} \langle \tau \rangle$; each period lasts 1600 time units. Note large transitions in $Q$ ($m$ changes sign). (F) $t_{1/2} = 4000 \gg \langle \tau \rangle$; note limited scale of ordinate.

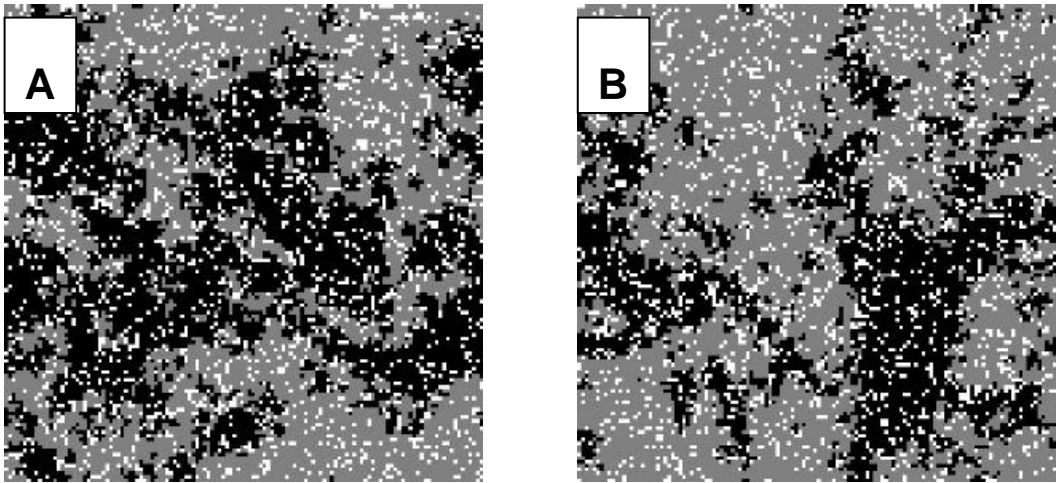

**Fig. 4**. MC-regime, $t_{1/2} = 100 << \langle \tau \rangle$. Open: white, Species 1: gray, Species 2: black. (A) Lattice at end of half-period during which species 2 had propagation advantage. (B) Lattice at end of next half-period; species 1 had propagation advantage. Neither species establishes large-scale spatial order; competitors coexist with densities fluctuating.

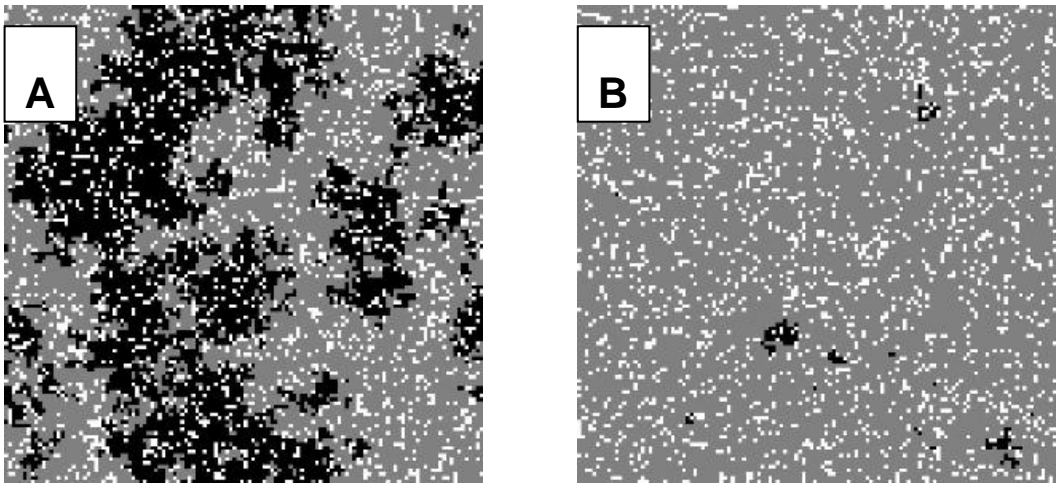

**Fig. 5**. MC-regime, $t_{1/2} = 800 \stackrel{\approx}{<} \langle \tau \rangle$. Open: white, Species 1: gray, Species 2: black. (A) Lattice at end of half-period during which species 2 had propagation advantage; $m(t) \approx 0$. (B) Lattice at end of next half-period; species 1 had propagation advantage; $m(t) \approx 0.9$ at completion of invasion-exclusion cycle.

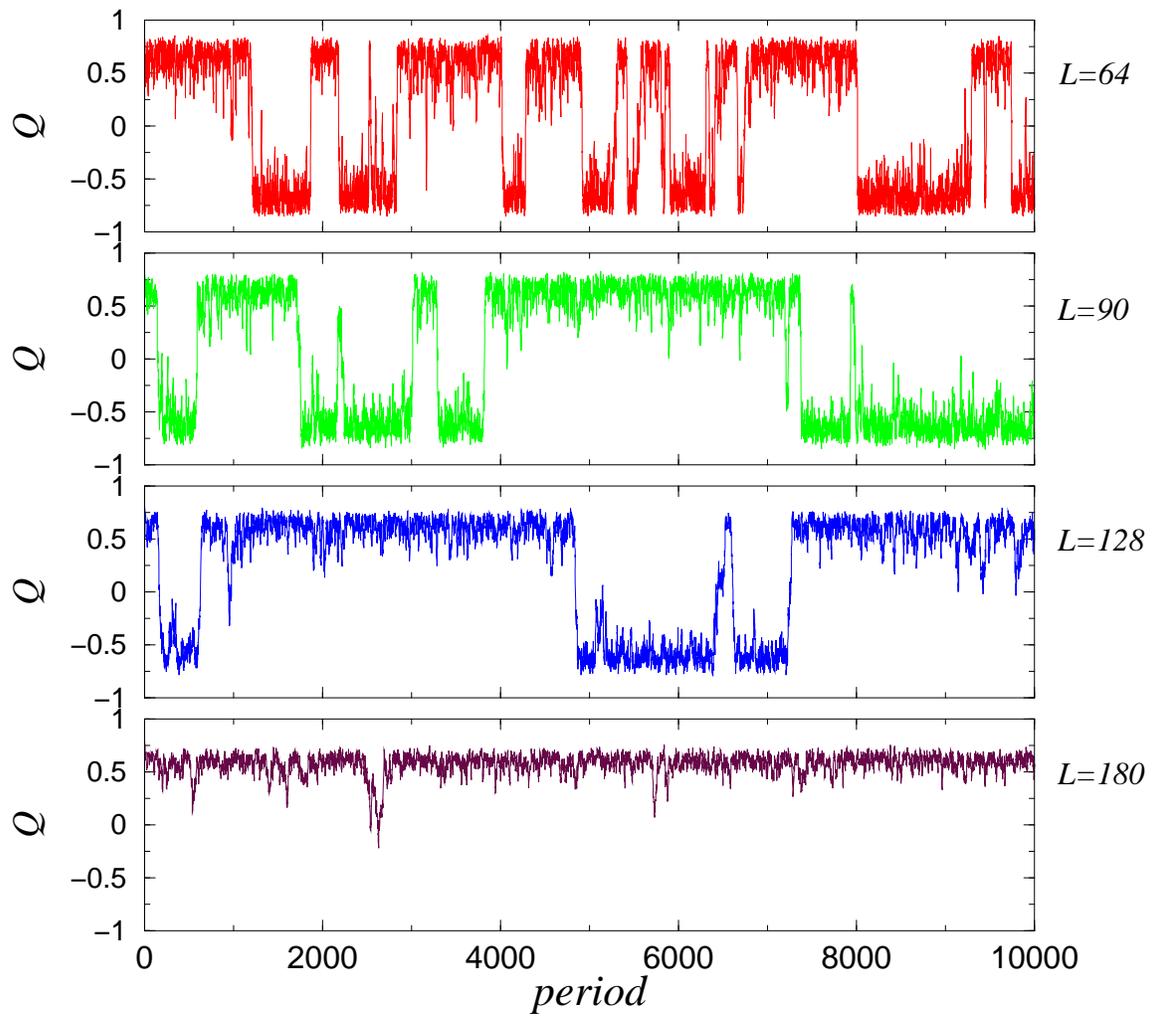

**Fig. 6.** Spontaneous dynamic symmetry breaking in the MC regime for $t_{1/2} = 800 \stackrel{\approx}{<} \langle \tau \rangle$. Period-averaged density difference $Q$ for $10^4$ consecutive periods for four habitat sizes $L = 64$, 90, 128, 180 (from top to bottom) for the same parameter values as in Fig. 3E.

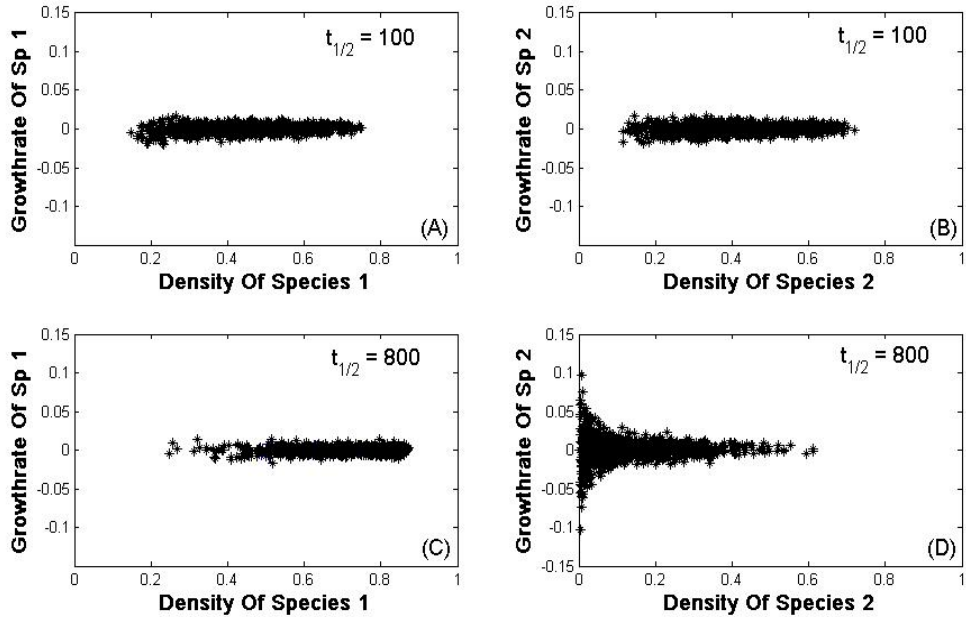

**Fig. 7**. Logarithmic growth rates, as functions of density, MC-regime. Species 1 (plots A, C) is resident, with initial density $\approx 0.86$. Species 2 (plots B, D) is the invader, with initial density 0. A, B: $t_{1/2} = 100 << \langle \tau \rangle$; C, D: $t_{1/2} = 800 \stackrel{\approx}{<} \langle \tau \rangle$. Each plot includes 1000 estimates, sampled regularly at intervals of 500 time units. When the environment oscillates relatively rapidly (plots A, B) non-equilibrium coexistence allows each species to traverse the same range of global densities over $10^5$ time units. The growth-rate plots are nearly identical. When timescales match (plots C, D) we observed the invader more often at low density, a consequence of initial conditions. Furthermore, matching timescales ($t_{1/2} = 800$) indicate invasion-exclusion cycling in invader growth; growth-rate variability is relatively large for $\rho_2(t) < 0.2$.

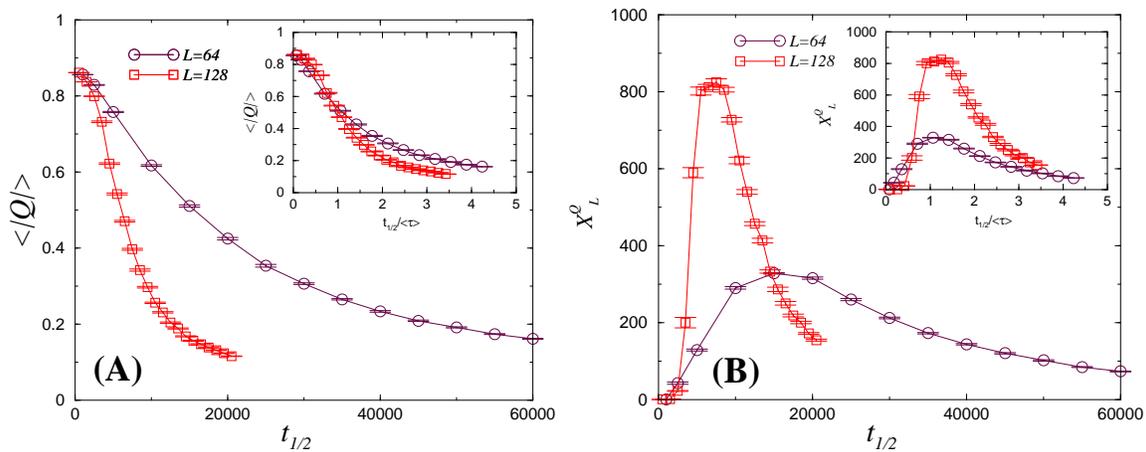

**Fig. 8**. Metrics for single-cluster invasion. (A) Average of absolute value of $Q$ as function of environmental half-period $t_{1/2}$. $\beta = 10^{-6}$; habitat size $L = 64, 128$; $\langle \tau \rangle \approx 14000, 6000$, respectively. (B) The scaled variance of $Q$. Single peak indicates value of $t_{1/2}$ where transition in competitive dynamics occurs. Inset: Rescaling shows singular phenomenon.

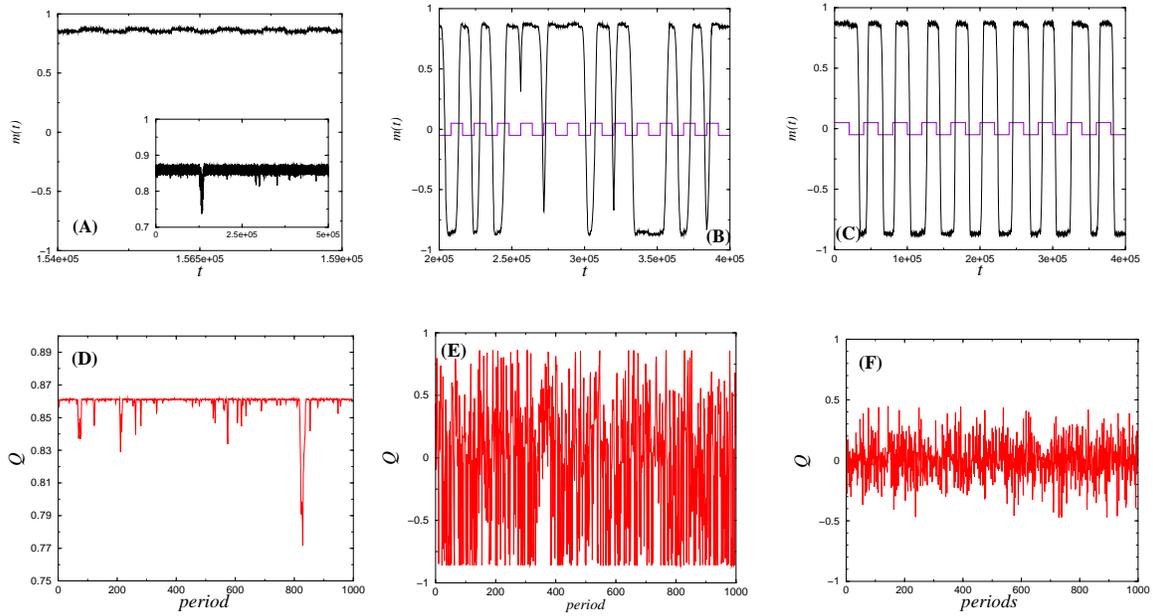

**Fig. 9**. Single-cluster regime: segments of the density difference $m(t)$ as a function of time, and period-averaged density difference $Q$ plotted for $10^3$ consecutive periods. $\beta = 10^{-6}$; habitat size $L = 128$; $\langle \tau \rangle \approx 6000$. (A) $t_{1/2} = 400 << \langle \tau \rangle$. Note limited scale of ordinate. Inset: Long-term plot, showing that species 2 never advanced significantly during simulation. (B) $t_{1/2} = 8000 \approx \langle \tau \rangle$. Stochastic resonance. (C) $t_{1/2} = 20000 >> \langle \tau \rangle$. System reaches single-species equilibrium during each, lengthy half-period. (D) $t_{1/2} = 400 << \langle \tau \rangle$; each period lasts 800 time units. Note limited scale of ordinate. (E) $t_{1/2} = 8000 \approx \langle \tau \rangle$; each period lasts 16000 time units. (F) $t_{1/2} = 20000 >> \langle \tau \rangle$; *cf*. Fig. 3F.

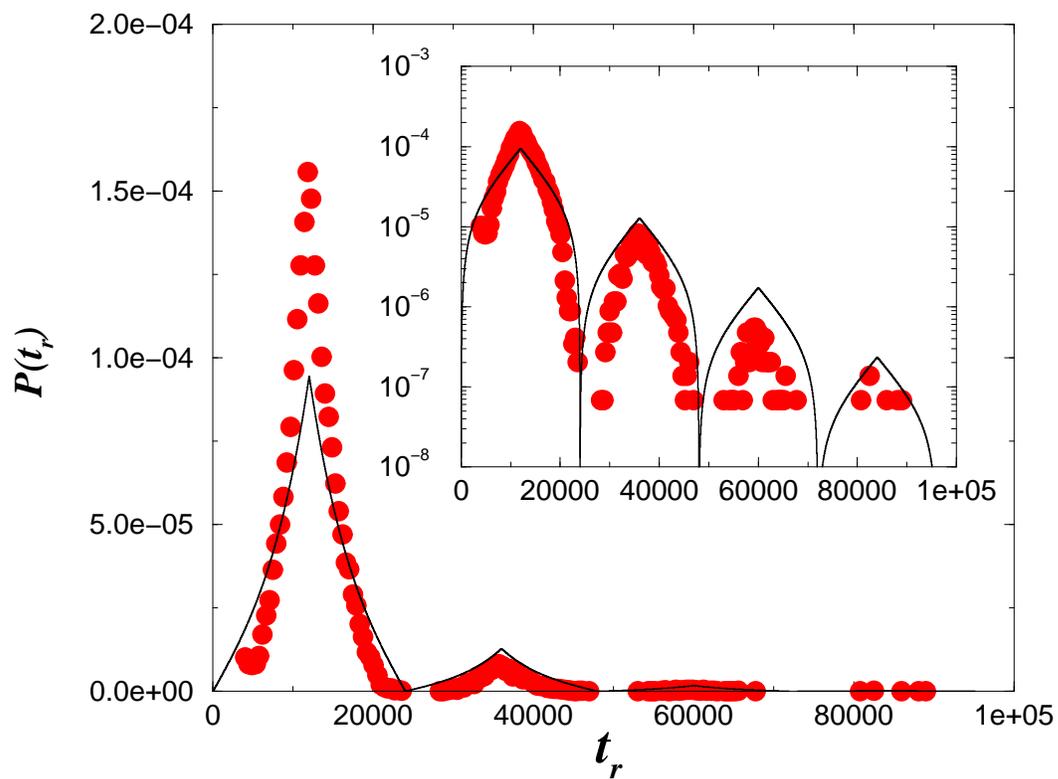

**Fig. 10**. Waiting-time distribution under stochastic resonance. Abscissa is waiting time between consecutive zero-crossings of $m(t)$; $\beta = 10^{-6}$, $L = 128$, $t_{1/2} = 1.2 \times 10^4$. Solid curves: theoretical density; small circles: observed waiting times, which peak at odd multiples of half-period - a signature of stochastic resonance.

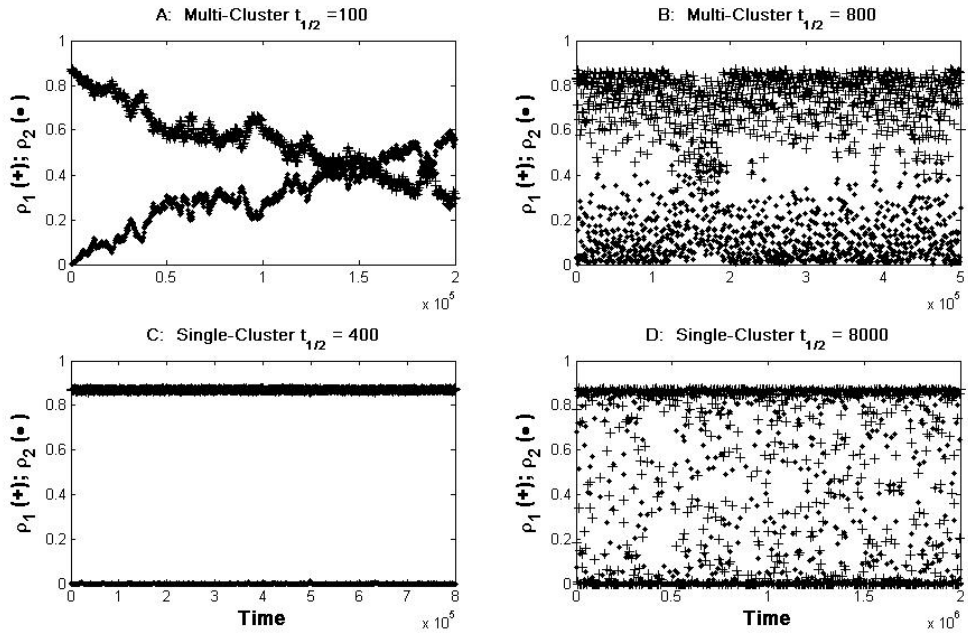

**Fig. 11**. Time-dependent densities. $L = 128$. Initial densities: $[\rho_1(t), \rho_2(t)] = (\rho_1^*, 0)$. (A) MC-regime; $t_{1/2} = 100 << \langle \tau \rangle$, non-equilibrium coexistence. (B) MC-regime; $t_{1/2} = 800 \overset{\approx}{<} \langle \tau \rangle$, invasion-exclusion cycles. (C) SC-regime; $t_{1/2} = 400 << \langle \tau \rangle$. No invasion observed. (D) SC-regime; $t_{1/2} = 8000 \approx \langle \tau \rangle$. Stochastic resonance.

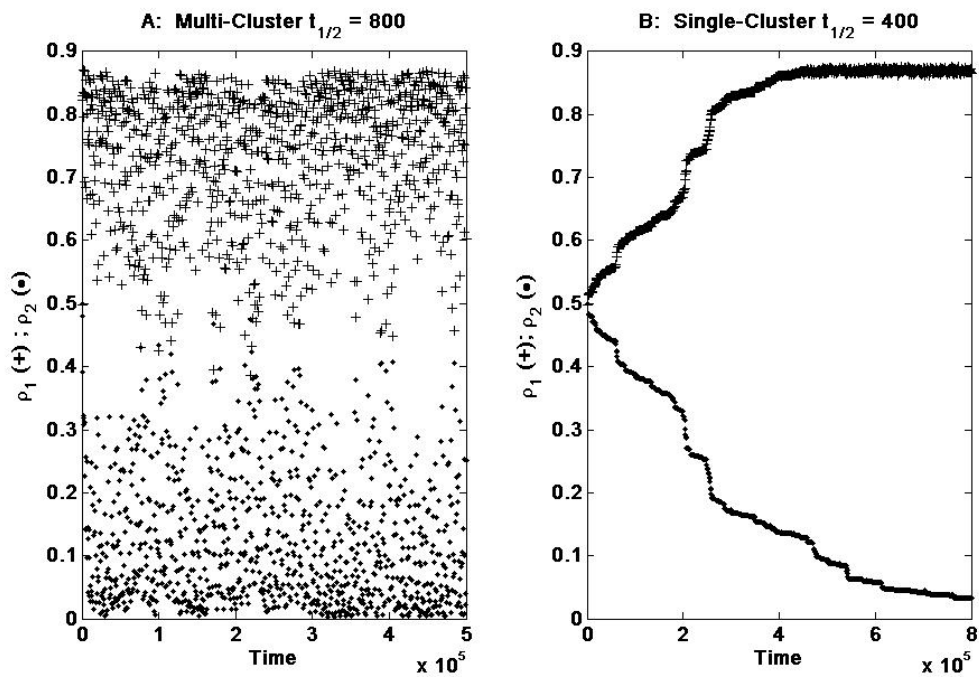

**Fig. 12.** Time-dependent densities. $L = 128$. Initial densities: $[\rho_1(t), \rho_2(t)] = (0.5, 0.48)$.
(A) MC-regime $t_{1/2} = 800 \overset{\approx}{<} \langle \tau \rangle$; Invasion-exclusion cycles, as in Fig. 11B. (B) SC-regime; $t_{1/2} = 400 << \langle \tau \rangle$. System attracted to single-species equilibrium as in Fig. 11C.